\documentclass[a4paper, 11pt]{article}
\usepackage[affil-it]{authblk} 
\usepackage{etoolbox}
\usepackage{lmodern}
\usepackage[a4paper]{geometry}
\usepackage[in]{fullpage}
\usepackage[utf8]{inputenc}
\usepackage{amsmath}
\usepackage{amssymb}
\usepackage{cite}

\usepackage{amsfonts}
\usepackage{graphicx}
\usepackage[dvipsnames]{xcolor}
\usepackage[colorlinks]{hyperref}
\hypersetup{linkcolor=blue,citecolor=blue,urlcolor=blue}

\title{Non-linear stability of \ensuremath{\alpha'}-corrected Friedmann equations}

\makeatletter
\patchcmd{\@maketitle}{\LARGE \@title}{\fontsize{16}{19.2}\selectfont\@title}{}{}
\makeatother

\author[1]{Heliudson Bernardo\footnote{Email: \href{mailto:heliudson.deoliveirabernardo@mgcill.ca}{heliudson.deoliveirabernardo@mcgill.ca}}}

\author[2]{Jan Chojnacki\footnote{Email: \href{matilto:jr.chojnacki@uw.edu.pl}{jr.chojnacki@uw.edu.pl}}}

\author[1]{Vincent Comeau\footnote{Email: \href{mailto:vincent.comeau@mail.mcgill.ca}{vincent.comeau@mail.mcgill.ca}}}

\affil[1]{Department of Physics, McGill University, Montreal, QC, H3A 2T8, Canada}

\affil[2]{Faculty of Physics, University of Warsaw,
Pasteura 5, Warsaw, Poland}

\date{\vspace{-5ex}}

\begin{document}

\maketitle


\begin{abstract}
We study the non-linear stability of fixed-point solutions to the \ensuremath{\alpha'}-exact equations from O\ensuremath{(d,d)} invariant cosmology, with and without matter perturbations. Previous non-linear analysis in the literature is revisited, and its compatibility with known linear perturbation results is shown. Some formal aspects of cosmological perturbations in duality invariant cosmology are discussed, and we show the existence of time-reparameterization invariant variables for perturbations.
\end{abstract}



\section{Introduction}
\label{sec:intro}
A robust prediction of all low-energy superstring effective ten-dimensional actions is their massless spectrum, which contains a spin-2 graviton,  a scalar dilaton field, and an anti-symmetric 2-form field. The string dynamics of these fields might be relevant when considering high temperatures while approaching the Big-Bang singularity \cite{Brandenberger:1988aj, Battefeld:2005av}. A remarkable feature of the string-theoretic low-energy effective action, absent in other gravitational theories, is a global, continuous O$(d,d)$ symmetry discovered in \cite{Meissner:1991zj}, and developed later in \cite{Tseytlin:1991wr,Sen:1991zi,Maharana:1992my,Kaloper_1997}. The symmetry is present on backgrounds with $d$ Abelian isometries, such as when the fields do not explicitly depend on $d$ out of $D$ coordinates, as in cosmology. This duality is currently in its renaissance, and new developments appear each year \cite{Hohm:2019ccp,Hohm:2019jgu,Wang:2019mwi,Krishnan:2019mkv,Wang:2019kez,Wang:2019dcj,Bernardo:2019bkz,Nunez:2020hxx,Codina:2020kvj,Basile:2021euh,Basile:2021krk,Bernardo:2021xtr,Quintin:2021eup,Codina:2021cxh,Codina:2022onm}.

The low-energy effective action describing the massless string gravity sector is given by
\begin{equation}\label{EffectiveAction}
S=\frac{1}{2\kappa^2}\int d^{d+1} x\, \sqrt{-G} e^{-2\phi}\left(R+4G^{\mu\nu}\partial_\mu \phi \partial_\nu \phi 
    -\frac{1}{12}H_{\mu\nu\rho} H^{\mu\nu\rho} \right),
\end{equation}
where $d$ is the number of spatial dimensions, $R$ is Ricci scalar associated to the metric $G$, and $H_{\mu\nu\rho}$ are the components of the field strength $H_3 = dB_2$ associated to the Kalb-Ramond field $B_{\mu\nu}$. For cosmological backgrounds, the fields ansatze are $G_{00}=-n^2(t),\, G_{0i}=0,\, G_{ij}=g_{ij}(t)$, $B_{00}=0,\, B_{0i}=0,\, B_{ij}=b_{ij}(t)$,  $\phi=\phi(t)$, and the above effective action may be expressed in the manifestly O$(d,d)$-invariant form \cite{Meissner:1991zj}
\begin{equation}\label{invariant action}
    S=\frac{1}{2\kappa^2}\int dt\,n(t) e^{-\Phi} \left[-(\mathcal{D}\Phi)^2-\frac{1}{8} \textrm{tr}\left(\mathcal{D}\mathcal{S}\right)^2 \right],
\end{equation}
with $\mathcal{S}$ defined as
\begin{equation}\label{eq: S definition}
\mathcal{S} = \begin{pmatrix}
			bg^{-1} & g - bg^{-1}b\\
				g^{-1} & -g^{-1}b
			\end{pmatrix},\\  
\end{equation}
and $\mathcal{D} = n^{-1} \partial_t$. The generalized dilaton 
$\Phi = 2\phi - \ln \sqrt{\det g_{ij}}$  and the lapse $n(t)$ transform as scalars under the O$(d,d)$ group, while $\mathcal{S}$ is an O$(d,d)$ tensor,
\begin{equation}\label{eq:Odd transformation}
    n \to n,\quad\Phi \to \Phi,\quad \mathcal{S}\to\Omega^T \mathcal{S}\Omega, \quad \text{where} \quad \Omega \in O(d,d).
\end{equation}
This is a generalized version of the scale-factor duality \cite{Meissner:1991zj}, where the symmetry group is discrete, and a cosmological scale factor transforms to its inverse. 

The first $\alpha'$ correction to the low-energy effective action can also be cast into  an O$(d,d)$-invariant form \cite{Meissner1997}. Moreover, string field theory implies that the duality symmetry should be present to all orders in $\alpha'$ \cite{Sen:1991zi}. The manifest O$(d,d)$ invariant formulation for cosmology became particularly interesting since the seminal paper \cite{Hohm:2019jgu}, where all possible $\alpha'$ corrections to the effective action were classified, providing non-perturbative solutions to the resulting $\alpha'$-exact Friedmann equations. Such equations may be trusted to a much higher energy scale than most previously known string corrections to General Relativity (GR). For an isotropic and homogeneous FLRW ansatz, a general action including all perturbative $\alpha'$-corrections and invariant under global O$(d,d)$ transformation \eqref{eq:Odd transformation} can be cast into the form \cite{Hohm:2019jgu, Bernardo:2019bkz}
\begin{equation}\label{eq:hohmzwiebachaction}
		S =\frac{1}{2\kappa^2} \int\,dt\,n(t)e^{-\Phi} \left[-(\mathcal{D}\Phi)^2 + \sum_{k=1}^{\infty} \alpha'^{k-1}c_k \mathrm{tr}\left(\mathcal{D}\mathcal{S}\right)^{2k} \right] + S_{\text{m}},
\end{equation}
with $c_1 = -1/8$. The coupling constants $c_{k>1}$ are generally unknown but can, in principle, be determined perturbatively from the worldsheet anomaly cancellation \cite{Bonezzi:2021mub,Bonezzi:2021sih}. We have also included matter sources with an action $S_{\text{m}}$, which is relevant for discussing realistic cosmological models \cite{Bernardo:2019bkz}. Equations of motion following such a non-perturbative action could provide novel insights into the physics of the early universe. The consequences of this non-perturbative completion have been intensively studied in the hope of finding unique observable signatures \cite{Bernardo:2020nol, Bernardo:2020bpa}.

For FLRW spacetimes, we can explicitly write the trace in \eqref{eq:hohmzwiebachaction} in terms of the Hubble rate $H= \mathcal{D}\ln a$. Doing so, we have 
\begin{equation}\label{eq:hohmzwiebachaction2}
	S =\frac{1}{2\kappa^2} \int\,dt\,n(t)e^{-\Phi} \left[-(\mathcal{D}\Phi)^2 - F(H) \right] + S_{\text{m}},
\end{equation}
where 
\begin{equation}\label{eq:F_function}
    F(H) = 2d \sum_{k=1}^{\infty}(-\alpha')^{k-1}c_k 2^{2k}H^{2k}
\end{equation}
encodes all $\alpha'$ corrections. Varying the action \eqref{eq:hohmzwiebachaction2} with respect to $n(t)$, $a(t)$ and $\Phi(t)$ gives the respective equations of motion:
\begin{subequations}\label{dualityinveqs}
\begin{align}
    (\mathcal{D}\Phi)^2 + H F'(H) - F(H) &= 2\kappa^2 e^{\Phi}\Bar{\rho},\\
    (\mathcal{D}H) F''(H) - (\mathcal{D}\Phi) F'(H) &= -2d\kappa^2 e^{\Phi}\Bar{p}, \label{scalefactoreq}\\
    2\mathcal{D}^2\Phi - (\mathcal{D}\Phi)^2 +F(H) &= \kappa^2e^{\Phi}\Bar{\sigma}.
\end{align}
\end{subequations}
The matter sources on the right-hand sides come from varying the matter action with respect to the fields,
\begin{equation}
    \rho = -\frac{2}{\sqrt{-G}}\frac{\delta S_{\text{m}}}{\delta g^{00}}, \quad p = -\frac{2}{d\sqrt{-G}}g^{ij}\frac{\delta S_{\text{m}}}{\delta g^{ij}},\quad \sigma = -\frac{2}{\sqrt{-G}} \frac{\delta S_{\text{m}}}{\delta \Phi},
\end{equation}
and the bar over those denote multiplication by $\sqrt{g}$, e.g. $\Bar{\rho} = \sqrt{g}\rho$. One could also vary \eqref{eq:hohmzwiebachaction} to get a matrix equation for $\mathcal{S}$ sourced by an O$(d,d)$ energy-momentum tensor, the trace of which gives \eqref{scalefactoreq}. See \cite{Gasperini:1991ak, Bernardo:2019bkz} for more details on matter coupling in the duality invariant setup. 

Note that the O$(d,d)$ group symmetry is spontaneously broken into the discrete symmetry
\begin{equation}
    n\to n, \quad \Phi \to \Phi, \quad a\to \frac{1}{a}, \quad \bar{\rho} \to \bar{\rho}, \quad \bar{p}\to -\bar{p}, \quad \Bar{\sigma} \to \Bar{\sigma},
\end{equation}
as manifest in \eqref{dualityinveqs} due to the even parity of $F(H)$.

De Sitter (dS) solutions for \eqref{dualityinveqs} are present in the non-perturbative regime of the duality-invariant cosmology \cite{Hohm:2019jgu, Bernardo:2019bkz, Bernardo:2020zlc}. Due to the lack of knowledge of the coefficients $c_k$ in \eqref{eq:F_function}, these solutions play a crucial role because, since $H=H_0$ is constant for them, $F(H)$ and its derivatives reduce to constant factors in the equations of motion. Moreover, since the generalized dilaton $\Phi$ can still be non-trivial, the solutions might not be dS in the Einstein frame, and we can have a non-trivial cosmological evolution in that frame, notwithstanding the functional form of $F(H)$. Such solutions were used to construct a background for string gas cosmology \cite{Brandenberger:1988aj} in \cite{Bernardo:2020nol}.

For the vacuum case, it was shown in \cite{Hohm:2019jgu} that the exponential growth of the scale factor is possible only when all $\alpha'$ corrections are 
included in the equations of motion. The linear stability of these and other solutions in the presence of matter was studied in \cite{Bernardo:2020zlc} (see also \cite{Nunez:2020hxx} for solutions including the $b$-field), with a special emphasis on the matter-coupled case since such solutions were used in the semi-realistic model of \cite{Bernardo:2020nol}. For vacuum dS solutions, the conclusion is that stability depends on the sign of the generalized dilaton's velocity. Stable solutions correspond to $\dot\Phi<0$, while unstable solutions have $\dot\Phi>0$. The special case $\dot\Phi=0$ required a higher-order stability analysis, which was not considered in \cite{Bernardo:2020zlc}. See also \cite{Quintin:2018loc} for the stability analysis of a matter-sourced solution including the leading $\alpha'$ correction.

In section \ref{sec:nonlinearstability}, we study the non-linear stability of the Minkowski and dS solutions for vanishing dilaton velocity case, $\dot{\Phi} =0$, with and without matter perturbations. For the vacuum dS case, the non-linear stability analysis was performed in \cite{Bieniek:2022mrv}, where it was found that these dS solutions are unstable. We revisit this case in section \ref{sec:anotherapproach}, where we show that while we agree with the results in \cite{Bieniek:2022mrv}, their approach cannot fully recover the linear stability results of \cite{Bernardo:2020zlc} when $\dot\Phi \neq0$. We explain why this is the case, making both approaches compatible. 
We also discuss the effect of matter perturbations for dS and Minkowski solutions for which a non-linear stability analysis is required.

In section \ref{sec:Bardeenvariables}, we make the first step towards an alternative approach, investigating the interplay between O$(d,d)$ and the remaining diffeomorphism symmetry after assuming a cosmological ansatz, i.e., time-reparameterization invariance. The actions \eqref{eq:hohmzwiebachaction} and \eqref{eq:hohmzwiebachaction2} are manifestly time-reparameterization invariant. However, one might be interested in a gauge-invariant description of the cosmological perturbation variables, which are known not to be gauge-invariant. In GR, the Bardeen variables \cite{Bardeen:1980kt} provide such a description, which is useful for tracking the evolution of perturbation modes beyond the Hubble radius in inflationary cosmology \cite{Mukhanov:1990me}. We show that the analog of such variables does exist in the duality invariant framework, even though the perturbations considered are much more restricted than in GR. A solution to these issues is particularly relevant if one wants to construct an inflationary model with the non-perturbative stable de Sitter solutions.

\section{Non-linear stability analysis}\label{sec:nonlinearstability}

\subsection{Review of linear fixed-point stability analysis}\label{sec:linearstability}

In this section, we recollect some results from \cite{Bernardo:2020zlc} about the stability analysis of fixed points of the $\alpha'$-corrected equations \eqref{dualityinveqs}. Setting $n(t)=1$, this system involves three equations for the variables $a(t),\, \Phi(t),\, \rho(t),\, p(t),\,$ and $\sigma(t)$. Solutions can be found if we assume constant equations of state, $p = w \rho$ and $\sigma = \lambda \rho$. Moreover, assuming $F''(H) \neq 0$ and denoting $\dot{\Phi} =y$, we can write equations \eqref{dualityinveqs} as
\begin{subequations}\label{autonomoussystem}
    \begin{align}
        \dot{H} &=  \frac{1}{F''}\left[ y F' - dw(y^2 +HF'-F) \right], \label{eq:ODE1}\\
        \dot{y} &= \frac{y^2}{2} - \frac{F}{2} + \frac{\lambda}{4} (y^2 + HF'-F), \label{eq:ODE2}\\
        C(H, y, \rho) &\equiv y^2 + H F' - F - 2 \kappa^2 e^\Phi\bar{\rho} =0. \label{eq:constraint}
    \end{align}
\end{subequations}
This is a dynamical system for $H(t)$ and $y(t)$ subject to the constraint $C(H, y, \bar{\rho}) =0$. The fixed points correspond to solutions with $\dot{H} = 0 = \dot{y}$, where we assume that the functional form of $F(H)$ is compatible with the constraint. For matter-sourced solutions, we have $\rho \propto e^{-\Phi}$ where the proportionality constant necessarily vanishes if $H_0 = 0= y_0$.

The stability of \eqref{autonomoussystem} around the fixed points is studied after linearizing the system. The time evolution of linear perturbations depends on the sign of the eigenvalues $\alpha_{\pm}$ of the Jacobian matrix associated with the linear system \cite{Bernardo:2020zlc},
\begin{equation}\label{eigenvalues}
    \alpha_{\pm} = -\frac{H_0\beta}{2} - \frac{H_0}{2} \left(\beta + \frac{\beta \lambda}{2} + dw\right) \pm \frac{1}{2}\sqrt{H^2_0 \left[ \frac{\lambda F'_0}{H_0} + \left(dw + \frac{\beta \lambda}{2} \right)^2 \right] -  \frac{2F'_0}{F''_0} (F'_0 + 2dw\beta H_0)}\,,
\end{equation}
where the constant $\beta(w,\lambda)$ is defined by $y_0 = -\beta H_0$ and $F''_0\equiv F''(H_0)$ is assumed to be non-vanishing. Positive eigenvalues characterize instabilities, and vanishing eigenvalues indicate that a non-linear analysis is needed to assess the system's stability. In \cite{Bernardo:2020zlc}, the sign of $\alpha_{\pm}$ was studied for different relevant choices of the equations of state, $w$ and $\lambda$. However, there are a few cases where $\alpha_{\pm} =0$, and a non-linear stability analysis becomes necessary, most notably for $\rho =0$ and no dilaton velocity, $\beta =0$. These cases include Minkowski and dS (in the string frame). Indeed, for the vacuum case, we should set $w =0 = \lambda$ in \eqref{autonomoussystem}, and eqs. \eqref{eq:ODE2} and \eqref{eq:constraint} imply $F_0'=0$, so that the eigenvalues degenerate to $\alpha_{\pm} = -\beta H_0 = y_0$. Thus, the solution is unstable for positive dilaton velocity, while the vanishing velocity or Minkowski cases require a higher-order stability analysis. Recently, in \cite{Bieniek:2022mrv}, the non-linear stability analysis for vacuum dS was carried out, and it was found that the $y_0=0$ case is unstable. However, their approach also implies that the $y_0\neq0$ vacuum solution is also unstable, which is clearly not true for $y_0<0$. In section \ref{sec:anotherapproach}, we revisit the incompatibility of these results, pointing out some potential issues in the methods employed in \cite{Bieniek:2022mrv}. 

\subsection{Stability analysis of Friedmann equations}

The non-linear stability analysis of \eqref{autonomoussystem} shares some similarities with the one which can be carried out for the ordinary Friedmann equations in general relativity. For an isotropic, homogeneous, and spatially flat spacetime containing a perfect fluid with density $\rho$ and pressure $p$, the Einstein equations imply that
\begin{subequations}
\begin{align}
    3H^2 &= \kappa^2_{\text{E}}\rho, \\
	\dot{H} &=-\frac12\kappa^2_{\text{E}}(\rho + p), 
\end{align} 
\end{subequations}
where $\kappa^2_{\text{E}} = 8\pi G_N$, and $G_N$ represents Newton's gravitational constant. Assuming a barotropic equation of state, $p = w\rho$, the second Friedmann equation then becomes
\begin{equation}
    \dot{H} = -\frac32 (1 + w) H^2.
\end{equation}
There exist two circumstances for which the Hubble rate is constant, $H = H_0$, and so $\dot{H} = 0$. First, it can be zero, $H_0 = 0$, which implies that the fluid density must also vanish $\rho_0 = 0$. The Hubble rate can also differ from zero, but the fluid must have an equation of state $w=-1$. These two cases, which correspond to the Minkowski and the de Sitter spacetimes, respectively, represent the fixed points of the Friedmann equations in standard cosmology.

Let's assess the stability of these fixed points. The proper way to do so would be to return to the Einstein equations and expand them around each fixed point by including general metric and matter perturbations. This approach is available in standard cosmology \cite{Mukhanov:1990me}, but not for the $\alpha'$-corrected Friedmann equations \eqref{dualityinveqs}. Hence, the best we can do to compare these two setups is to perturb the Friedmann equations directly by including the most general perturbations permitted by these equations. Specifically, we will write the Hubble rate and the fluid density in terms of their value at the fixed point, plus small corrections, $H = H_0 + \delta H(t), \rho = \rho_0 + \delta \rho(t)$. Here, we will not restrict to the linear analysis as in \cite{Bernardo:2020zlc}.

The de Sitter case is the most straightforward one. Assuming that the fluid retains the same equation of state $w=-1$, the second Friedmann equation then implies that $\dot{H} = 0$, which is true not only at the fixed point but also for the perturbations as well, at any order $\delta\dot{H} = 0$. Hence, the de Sitter spacetime is stable, at least in the context of the present analysis. 

The case of Minkowski spacetime is more interesting. With $H_0 = 0$ and $\rho_0 = 0$ at the fixed point, the perturbed Friedmann equations then become
\begin{subequations}
\begin{align}
    (\delta H)^2 &= \frac13 \kappa^2_\text{E} \delta \rho, \\
	\delta \dot{H} &= -\frac32 (1 + w) (\delta H)^2.
\end{align}
\end{subequations}
A purely linear analysis of the equations is  insufficient to assess the stability of the Minkowski spacetime since $\delta\dot{H} \simeq 0$ at the first order. The correction to the Hubble rate is then a mere constant $\delta H \simeq \delta A$, related to the perturbation in the fluid's energy density, but it becomes a source term for the second-order correction. After integrating, we get
\begin{equation}
    \delta H \simeq \delta A - \frac32 (1 + w) (\delta A)^2 \,t.
\end{equation}
Thus, when perturbed at the second order, the Hubble rate of a Minkowski spacetime ceases to be zero and acquires a time dependence that varies linearly with the physical time. This contribution to the Hubble rate is either positive or negative depending on the equation of state. For $w>-1$, it eventually becomes negative, whatever the sign of the constant $\delta A$.

Moreover, the correction we just derived does not vanish because we have included matter perturbations in our analysis. If instead, we do not allow matter perturbations, $\delta \rho = 0$, then $\delta H = 0$, and the Minkowski spacetime appears stable. On the other hand, if matter perturbations are included, it produces a gravitational instability and eventually causes the spacetime to contract when $\delta H$ becomes negative. In that case, the Minkowski spacetime represents an unstable fixed point of the Friedmann equations. Its instability is due to non-linear effects and grows quite mildly, not exponentially, with time.  

In the next two sections, we conduct an analogous stability analysis for Minkowski and dS solutions of the $\alpha'$-corrected equations \eqref{autonomoussystem}.

\subsection{Stability of Minkowski solutions}

Let's first consider the case where the fixed point corresponds to Minkowski spacetime, $H_0 = 0$ and $\rho_0 = 0$. Since $F(0) = F'(0) = 0$, we must also have $y_0 = 0$ for equations \eqref{autonomoussystem} to be satisfied. Note that the solution in the Einstein frame is also a Minkowski spacetime. If the Hubble rate, the dilaton and the fluid energy density are all slightly perturbed,
\begin{align*}
    H &= H_0 + \delta H(t), \\[4pt]
	\Phi &= \Phi_0 + \delta \Phi(t), \\[4pt]
	\rho &= \rho_0 + \delta \rho(t),
\end{align*}
the $\alpha'$-corrected Friedmann equations then become, up to the second order, 
\begin{subequations}\label{eq:pert@Minkowski}
\begin{align}
    \kappa^2e^{\Phi_0}\delta\Bar{\rho}(1+\delta\Phi) &= \frac12(\delta y)^2 - \frac12 d (\delta H)^2, \\[5pt]
	\delta\dot{H} &= \delta H \delta y + w\kappa^2e^{\Phi_0}\delta\Bar{\rho}(1+\delta \Phi), \\[5pt]
	\delta\dot{y} &= \frac12 (\delta y)^2 + \frac12 d (\delta H)^2 + \frac\lambda2 \kappa^2e^{\Phi_0}\delta\Bar{\rho}(1+\delta \Phi).
\end{align}
\end{subequations}
To compare with our previous analysis for the standard Friedmann equations, let's first consider the case where no matter perturbations are allowed, $\delta\rho = 0$. In that case, one of the three Friedmann equations becomes redundant. The first and last equations imply that $\delta \dot{y} = (\delta y)^2$, which can be integrated as
\begin{equation}\label{y-minkowski}
    \delta y = \delta B + (\delta B)^2 t,
\end{equation}
where $\delta B$ is a constant. It follows that
\begin{equation}
	\delta\dot{H} = \delta B\, \delta H.
\end{equation}
At the first order, $\delta\dot{H} \simeq 0$, which is why a second-order analysis is necessary for this particular fixed point. After integrating, we get
\begin{equation}
    \delta H = \delta A + \delta A\, \delta B \, t.
\end{equation}
Here, $\delta A$ and $\delta B$ represent the constant perturbations induced at the first order. The perturbed Hubble rate is similar to the one we derived for the standard Friedmann equations, in the presence of matter perturbations. The shifted dilaton perturbation simulate ordinary matter perturbations, by causing a similar non-linear instability to the Minkowski spacetime in GR. It generates a second-order time dependence to its Hubble rate, which once again varies linearly with the physical time. This correction can be either positive or negative, depending on the sign chosen for the dilaton velocity (\ref{y-minkowski}), whereas it is  negative for the ordinary matter perturbations in standard cosmology (when $w>-1$). 

More generally, when matter perturbations are included, equations \eqref{eq:pert@Minkowski} can be combined as
\begin{subequations}
\begin{align}
    \delta\dot{H} &= \delta H \,\delta y - \frac12 w d\,(\delta H)^2 + \frac12 w (\delta y)^2, \\[5pt]
	\delta\dot{y} &= \frac12 d\left(1 - \frac\lambda2\right)(\delta H)^2 + \frac12\left(1 + \frac\lambda2\right)(\delta y)^2 .
\end{align}
\end{subequations}
Integrating up to the second order, we get
\begin{subequations}
\begin{align}
    \delta H &= \delta A + \left(\delta A \,\delta B - \frac12 w d\,(\delta A)^2 + \frac12 w (\delta B)^2 \right) t, \\[5pt] 
	\delta y &= \delta B + \left(\frac12 d(1 - \lambda)(\delta A)^2 + \frac12(1 + \lambda)(\delta B)^2 \right) t,
\end{align}
\end{subequations}
where $\delta A$ and $\delta B$ are the constant perturbations induced at the first order. Once again, the Hubble rate and the dilaton velocity are modified by terms which vary linearly with time, thus causing a non-linear instability of the Minkowski spacetime.

\subsection{Stability of dS solutions}

Let's now consider the case where the fixed point corresponds to de Sitter spacetime, with a constant expansion rate $H_0 \neq 0$. The stability of such a fixed point has been already assessed in \cite{Bernardo:2020zlc}, except in the special case when the dilaton velocity vanishes, $y_0 = 0$, for which a purely linear analysis is not sufficient as studied in \cite{Bieniek:2022mrv}. 

For the Friedmann equations to be satisfied when $y_0 = 0 = \rho$ and $H_0\neq0$, the function $F$ must be such that $F(H_0) = 0$ and $F'(H_0) = 0$. Near the fixed point, this function must therefore be of the form
\begin{equation}
    F(H) = \frac12 F_0'' (H - H_0)^2 + \mathcal{O}((H - H_0)^3),
\end{equation}
where $F_0'' = F''(H_0) \neq 0$. If the Hubble rate, the dilaton velocity, and the fluid density are all slightly perturbed, the Friedmann equations then become
\begin{subequations}
\begin{align}\label{friedmann-de-sitter}
	\kappa^2e^{\Phi_0}\delta\Bar{\rho}(1+\delta\Phi) &= \frac12 H_0 F_0'' \delta H + \frac12 (\delta y)^2, \\[5pt]
	\delta\dot{H} &= -\frac{2wd}{F_0''}\,\kappa^2e^{\Phi_0}\delta\Bar{\rho}(1+\delta\Phi) + \delta H \,\delta y, \\[5pt]
	\delta\dot{y} &= \frac12 \lambda\kappa^2e^{\Phi_0}\delta\Bar{\rho}(1+\delta \Phi) + \frac12 (\delta y)^2.
\end{align}
\end{subequations}
Here, we only kept the terms linear in the perturbations $\delta H$ and $\delta\rho$, as well as those at most quadratic in $\delta y$, for the following reason: in the absence of matter perturbations, $\delta\rho = 0$, the first of these equations implies that
\begin{equation}\label{H-de-sitter}
	\delta H = -\frac{1}{H_0 F_0''}\,(\delta y)^2.
\end{equation}
In other words, when matter perturbations are not allowed, the correction to the Hubble rate is necessarily of the second order. This is why the terms quadratic in $\delta H$ have been dropped in the perturbed Friedmann equations. Moreover, when $\delta\rho = 0$, the evolution equation for the dilaton velocity is
\begin{equation}\label{eqfordeltayalone}
    \delta\dot{y} = \frac12 (\delta y)^2,
\end{equation}
that is, after integrating,
\begin{equation}\label{solfordeltayalone}
	\delta y = \delta B + \frac12 (\delta B)^2\, t,
\end{equation}
where $\delta B$ is the constant perturbation induced at the first order. The correction to the Hubble rate is then
\begin{equation}
	\delta H = -\frac{1}{H_0 F_0''} (\delta B)^2 \big(1 + \delta B\,t\big).
\end{equation} 
Thus, as for Minkowski spacetime, de Sitter spacetime also exhibits a non-linear instability induced by the string corrections. Its expansion rate acquires a linear time-dependence and so does the dilaton velocity.

Such an instability also occurs when matter perturbations are included, although it could seem like the latter tend to make the spacetime stable, when considered in a purely linear analysis. Indeed, when $\delta\rho\neq0$, the correction to the Hubble rate no longer vanishes at the first order. Neglecting all terms quadratic in the perturbations in the Friedmann equations (\ref{friedmann-de-sitter}), we are left with
\begin{subequations}
\begin{align}
    \kappa^2e^{\Phi_0}\delta\bar{\rho} &= \frac12 H_0 F_0''\, \delta H, \\[5pt]
	\delta\dot{H} &= -wd H_0\,\delta H , \\[5pt]
	\delta\dot{y} &= \frac14 \lambda H_0 F_0'' \,\delta H . 
\end{align}
\end{subequations}
Integrating the two evolution equations, we get
\begin{subequations}
\begin{align}
    \delta H &= \delta A \,e^{-wdH_0t}, \\[5pt]
	\delta y &= -\frac{\lambda F''_0}{4wd}\,\delta A\,e^{-wdH_0t} + \delta B,
\end{align}
\end{subequations}
where $\delta A$ and $\delta B$ are the two constants of integration. The linear analysis in \cite{Bernardo:2020zlc} does indeed predict the two eigenvalues $\alpha_{+} = 0$ and $\alpha_{-} = -wdH_0$, in the special case when $y_0 = 0$ and $F_0' = 0$ (see eq. \eqref{eigenvalues}). 

When $w<0$, the previous solutions grow exponentially with time, and the fixed point is thus unstable. When $w>0$, none of the previous solutions contains an exponential with a positive exponent. We might then be tempted to conclude that the fixed point is stable. However, the linear correction to the dilaton velocity contains a constant term, $\delta B$, which does not decrease with time, and therefore contributes to the growth of the perturbations at the second order. Taking $\delta H \simeq 0$, $\delta \rho \simeq 0$, and $\delta y \simeq \delta B$ at the first order, we get expressions for the perturbations which include a second-order correction varying linearly with time, just as in our previous calculation, when matter perturbations were not allowed. Hence, the latter does not eliminate the non-linear instability of de Sitter spacetime for vanishing dilaton velocity. 

\subsection{Another approach to the non-linear stability analysis of dS solutions}\label{sec:anotherapproach}

The vacuum results of the previous section about the non-linear instability of de Sitter spacetime induced by the full string corrections had already been noticed in \cite{Bieniek:2022mrv}. We shall quickly review their approach, and discuss some potential issues regarding vacuum dS instabilities. The analysis of \cite{Bieniek:2022mrv} is based on manipulating the vacuum equations
\begin{subequations}
\begin{align}
    \dot{H} &= \frac{F'}{F''}y,\\
    \dot{y} &= -\frac{1}{2}H F',\\ \label{eq:vacuumconstraint}
    C(H,y) &= y^2 + HF' -F = 0, 
\end{align}
\end{subequations}
to find an equation of the form $\dot{H} = U(H)$, where $U(H)$ is function of $H$ only. This is achieved after using the constraint \eqref{eq:vacuumconstraint} to write $y$ as a function of $H$. Hence, by taking a time-derivative, the dynamics for $H(t)$ is fixed by an equation of the form $\ddot{H} = (U^2)'/2 \equiv W(H) $, which resembles the Newtonian dynamics of a particle subject to a non-trivial potential. Then, stability around a fixed point can be studied after expanding $W(H)$ around that solution:
\begin{equation}\label{fequalma}
    \delta \ddot{H} = W'(H_0)\delta H + \frac{1}{2}W''(H_0) (\delta H)^2 +\cdots \,.
\end{equation}
The function $W(H)$ is given by
\begin{align}
    W(H) = \frac{F'}{2
   (F'')^3} \left[(F'')^2\left(2F - 3HF'\right)+2F' F''' \left(HF' - F\right)\right],
\end{align}
and, for $y_0 = 0$, $W'(H_0)$ vanishes as expected from the discussion in section \ref{sec:linearstability}. Evaluating $W''(H_0)$ for $y_0 =0$ or, equivalently, taking two time derivatives of (\ref{H-de-sitter}), we get
\begin{equation}
	\delta\ddot{H} = -\frac32 H_0 F_0''\,(\delta H)^2.
\end{equation}
Hence, the behaviour of the perturbation $\delta H$ appears to be governed by a cubic potential. There must exist a direction along which $\delta H$ grows with time, which implies that de Sitter spacetime represents an unstable fixed point. Although this particular conclusion is confirmed by our previous analysis, the approach of \cite{Bieniek:2022mrv} we just described fails to give the stability of the $y_0< 0$ solution. To see that, consider the value of $W'(H_0)$ when $y_0\neq 0$,
\begin{equation}
    W'(H_0) = F_0 = y_0^2.
\end{equation}
According to \eqref{fequalma}, the solution is unstable regardless of the dilaton's velocity! Why does the method above yield the correct result for the non-linear perturbation case, but does not recover the linear stability analysis results? 

The answer is that the equation $\ddot{H} = W(H)$ carries less information about the system than the original equations; more specifically, $\delta\dot{H}$ is less constrained by it. For instance, let's assume that the perturbed Friedmann equation around a hypothetical fixed point takes the form
\begin{equation}
	\delta\dot{H} = \alpha_0\,\delta H,
\end{equation}
where $\alpha_0$ is a constant. Integrating, we get
\begin{equation}
	\delta H = \delta A\,e^{\alpha_0 t},
\end{equation}
and the fixed point is thus stable if $\alpha_0 < 0$, and unstable if $\alpha_0 > 0$. However, taking an additional time derivative as done in \cite{Bieniek:2022mrv}, we get alternatively $
    \delta\ddot{H} = \alpha_0\,\delta\dot{H} = \alpha_0^2\,\delta H,$ that is,
\begin{equation}
	\delta H = \delta A\,e^{|\alpha_0| t} + \delta B\,e^{-|\alpha_0| t}.
\end{equation}
Thus, the full solution to the second-order evolution equation for $\delta H$ not only includes the previous solution, with an exponent $\alpha_0$, but also another one with the opposite exponent. The fixed point then appears to be unstable for any value of $\alpha_0$, which is clearly an incorrect conclusion. We should therefore stick to the original first-order evolution equations for $\delta H$ and $\delta y$, and avoid taking additional time derivatives, when assessing the stability of the fixed points.

Moreover, there is already a potential issue with the equation $\dot{H} = U(H)$: the dynamics of system \eqref{autonomoussystem} cannot always be reduced to the dynamics of $H(t)$ because the function $U(H)$ might not be unique. To understand when this is so, let's take the derivative of the constraint $C(H,y)$,
\begin{equation}
    \dot{C} = 2y \dot{y} + HF'' \dot{H}=0,
\end{equation}
where the second equality follows from the dynamical equations \eqref{eq:ODE1} and \eqref{eq:ODE2}. So, $C(H,y)$ is a first-class constraint preserved during time evolution. This implies that we may forget about one of the evolution equations and consider only the remaining one plus the constraint, since the evolution for $H$ and $y$ are related. Note that such a simplification occurs only in vacuum. To solve the single dynamical equation, whichever one is chosen, we need to use the constraint to find an algebraic relation between $H$ and $y$. However, from the implicit function theorem \cite{spivak1965calculus}, we can solve $C(H,y) =0$ for $y$ and write a unique $y= y(H)$ near $(H_0, y_0)$ only if
\begin{equation}
    \left.\frac{\partial C}{\partial y}\right|_0 = 2y_0\neq 0.
\end{equation}
Therefore, writing $\dot{H} = U(H)$ is not appropriate around $y_0= 0$. This is precisely the case considered in \cite{Bieniek:2022mrv}. The reason why the non-linear stability results of \cite{Bieniek:2022mrv} in the $y_0=0=F_0''$ case are correct is that $W(H)$ is unique, although $U(H)$ is not.

On the other hand, since
\begin{equation}
    \left.\frac{\partial C}{\partial H}\right|_0 = H_0 F_0'' \neq 0
\end{equation}
for the relevant vacuum dS solutions, we can solve the constraint to write $H(y)$ around $(H_0, y_0)$, regardless of $y_0$. That's what was done in eq. \eqref{H-de-sitter}. After that, we found a dynamical equation for $\delta y$ alone (eq. \eqref{eqfordeltayalone}), and then solved it. In this way, possible technical inconsistencies were avoided.

For $F_0''=0 = y_0$, first discussed in \cite{Bieniek:2022mrv}, we cannot solve the constraint for $y$ or $H$. In this case, the perturbed equations are
\begin{subequations}
\begin{align}\label{firsteq}
    \delta\dot{H}\left(F_0'''+\frac{F_0^{(4)}}{2}\delta H\right) - \frac{F_0'''}{2}\delta y \delta H &=0, \\
    \delta \dot{y}- \frac{1}{2}(\delta y)^2 &= 0,\\ \label{thirdeq}
    (\delta y)^2 + \frac{1}{2}H_0 F'''_0 (\delta H)^2 &=0.
\end{align}
\end{subequations} 
The solution to the second equation is again \eqref{solfordeltayalone} and we can use this result to solve eq. \eqref{firsteq} to leading order
\begin{equation}
    \delta H = \delta A\left(1+ \frac{1}{2}\delta B t\right),
\end{equation}
where $\delta A$ is related to $\delta B$ through \eqref{thirdeq}. The fixed point is then unstable at non-linear order, as shown in \cite{Bieniek:2022mrv}.

\section{Comparison with standard cosmological perturbations: the existence of ``gauge-invariant'' variables}\label{sec:Bardeenvariables}

The perturbative analysis initiated in \cite{Bernardo:2020zlc} assumes specific types of perturbations. Since the structure of the $\alpha'$-corrections in the action \eqref{eq:hohmzwiebachaction2} requires flat FLRW metrics (because those have $d$ Abelian isometries), only scalar and homogeneous perturbations in $n(t)$ and $a(t)$ (and in the matter sector, when $S_\text{m}\neq0$) can be assumed. Although the results so far are self-consistent, we cannot turn on the vector and tensor perturbation modes as done in Einstein gravity, and stability under these more general perturbations cannot be assessed. 

Perturbations in the lapse function $n(t)$ are related to invariance under time reparameterization, since a change in $n(t)$ can be absorbed by redefining time. This redefinition, however, induces a change in the scale factor and so $\delta a$ is only uniquely defined once we specify the time variable. This is the usual redundancy associated with gauge symmetries manifesting itself in our theory: fixing $t$ is a choice of gauge. In standard cosmology, there are two ways of dealing with gauge symmetry under reparameterizations, one can fix a gauge and work out the perturbations in that gauge or one can define gauge invariant variables and write all the dynamical equations in terms of those. The former option has been used in \cite{Bernardo:2020zlc, Nunez:2020hxx, Bieniek:2022mrv}, where one fixes the gauge with $n=1$ and only perturb $a(t)$ and $\Phi(t)$ (and the matter variables in the sourced case). In this section, we show that it is possible to pursue the second option, i.e., work with time-reparameterization invariant variables in the context of $\alpha'$-complete cosmology.

To set some notation, let's again compare the $\alpha'$-complete setup to the standard theory for cosmological perturbations. Consider the action for Einstein gravity plus a scalar field,
\begin{equation}
    S =  \int d^4x\sqrt{-g}\left(\frac{1}{2\kappa^2_E} R - \frac{1}{2}g^{\mu\nu}\partial_\mu \phi \partial_\nu \phi -V(\phi)\right). 
\end{equation}
One finds an action for perturbations around solutions for this theory after defining small perturbations\footnote{In this section, we use subscripts and superscripts between parenthesis to denote the perturbation order of variables.} $\{g_{\mu\nu}^{(1)}, \phi^{(1)}\}$ around a solution $\{g^{(0)}_{\mu\nu}, \phi^{(0)}\}$, 
\begin{equation}
    g_{\mu\nu} = g_{\mu\nu}^{(0)} + g^{(1)}_{\mu\nu}, \quad \phi = \phi^{(0)} + \phi^{(1)}
\end{equation}
and expanding the action to quadratic order in the perturbations. The resulting theory for $\{g^{(1)}_{\mu\nu}, \phi^{(1)}\}$ has a local symmetry associated to infinitesimal coordinate transformations $x^\mu \to x^\mu + \xi^\mu$ that is generated by the Lie derivative
\begin{subequations}\label{infcoordtransf}
\begin{align}
    \delta_\xi g^{(1)}_{\mu\nu} &= \mathcal{L}_\xi g_{\mu\nu} = \xi^\rho \partial_\rho g_{\mu\nu}^{(0)} + g_{\nu \rho}^{(0)}\partial_\mu \xi^\rho + g_{\mu \rho}^{(0)}\partial_\nu \xi^\rho, \\
    \delta_\xi \phi^{(1)} &= \mathcal{L}_\xi \phi = \xi^\rho \partial_\rho \phi^{(0)}.
\end{align}
\end{subequations}
For FLRW solutions, the perturbations and gauge parameters are decomposed into scalar, vector and tensors of the $g_{ij}^{(0)}$ spatial isometry group. Then, gauge-invariant variables associated with the scalar modes are constructed, the so-called Bardeen variables \cite{Bardeen:1980kt} and the equations of motion are written with respect to those. We will not describe the whole procedure here, see \cite{Mukhanov:1990me} for reviews on cosmological perturbations and \cite{Chiaffrino:2020akd} for a modern view on the topic.  

Now, for
\begin{equation}\label{HZaction}
    S =\frac{1}{2\kappa^2} \int dt\, n(t) e^{-\Phi} \left[-(\mathcal{D}\Phi)^2 - F(H)\right],
\end{equation}
the time-reparameterization transformation of $n(t)$, $a(t)$ and $\Phi(t)$ can be obtained from \eqref{infcoordtransf} by imposing $\xi^\mu = (\xi^0, 0,\dots,0)$:
\begin{equation}\label{gaugetrans2}
    \delta_\xi g^{(1)}_{00} = 2 g_{00}^{(0)}\mathcal{D}\left(n\xi^0\right), \quad \delta_\xi g^{(1)}_{ij} = 2 n\xi^0 H_{(0)} g_{ij}^{(0)}, \quad \delta_\xi \Phi^{(1)} = n\xi^0 \mathcal{D}\Phi^{(0)},
\end{equation}
where the background metric has components $g_{00}^{(0)} = -n^2(t)$ and $g_{ij}^{(0)} = a^2(t)\delta_{ij}$ and we have used the definition $\Phi = 2\phi + \ln \sqrt{\det g_{ij}}$. Recall the definition of the derivative operator $ \mathcal{D}=n^{-1}\partial_t$ and that $H = \mathcal{D}\ln a$. From these transformation rules, it follows that $n^{(1)}$ transforms as a density while $a^{(1)}$ and $\Phi^{(1)}$ are scalars. 

In standard cosmological perturbation theory, the three-scalar gauge-invariant variables are constructed not only from $n^{(1)}$ and $a^{(1)}$ but also from the scalar modes $V$ and $T$ in the vector and tensor perturbations, $\partial_i V \subset g_{0i}^{(1)}$ and $(\partial_i\partial_j  -1/3\delta_{ij}\nabla^2) T \subset g_{ij}^{(1)}$, respectively. For the theory \eqref{HZaction}, we do not have these latter two modes, and one might ask whether it is possible or not to find time-reparameterization invariant variables. The answer turned out to be positive, due to the presence of the shifted dilaton, as shown below. 

We would like to find time-reparameterization invariant variables analogous to the Bardeen variables of cosmological perturbations. Since we take $\{n(t), a(t), \Phi(t)\}$ as independent variables and we have one degree of freedom in $\xi^0$, we are looking for two independent gauge-invariant variables $B_1$ and $B_2$ which are linear and homogeneous functions of the perturbations. It is straightforward to check that
\begin{subequations}
\begin{align}\label{eq:gaugeinv1}
    B_1 &= -\frac{1}{2d}\mathcal{D}\Phi^{(0)} g^{ij}_{(0)}g_{ij}^{(1)} + H_{(0)} \Phi^{(1)}\\ \label{eq:gaugeinv2}
    B_2 &= H_{(0)} \mathcal{D}\Phi^{(0)}g_{00}^{(1)}- 2 g_{00}^{(0)}H_{(0)} \mathcal{D}\Phi^{(1)} + \frac{1}{d} g_{00}^{(0)}\mathcal{D}^2 \Phi^{(0)} g^{ij}_{(0)}g_{ij}^{(1)},
\end{align}
\end{subequations}
satisfy these requirements. Hence, it is a priori possible to work with time-reparameterization invariant variables without the need for gauge fixing. 

Writing the perturbed equations of motion in terms of $B_1$ and $B_2$ will not be so straightforward. One reason is that in the discussion in section \ref{sec:nonlinearstability}, a perturbative expansion in terms of the Hubble expansion rate $H(t)$ is assumed instead of an expansion in terms of the scale factor (which is the fundamental metric variable). More importantly, expressing $H^{(1)}$ and $\Phi^{(1)}$ in terms of the gauge-invariant variables will necessarily introduce non-locality in time. To see this, first note that due to the density nature of $n^{(1)}(t)$, a term containing time-derivatives of the other perturbed fields is always needed in order to construct some gauge-invariant variable, as in \eqref{eq:gaugeinv2}. Hence, even after using the time-reparameterization symmetry to set $n^{(1)}=0$, we will not get an algebraic equation involving $\Phi^{(1)}(t)$, $a^{(1)}(t)$ and the $B$'s. This is in contrast with GR, in which we can choose the relation between the scalar perturbations in the metric and the Bardeen variables to be algebraic (in the conformal Newtonian gauge, for instance, one of the two scalar perturbations is equal to the first Bardeen variable while the other differs from the second by a sign).

Another feature worth commenting on is that, since \eqref{HZaction} does not depend on the spatial directions, the perturbation theory of the associated $\alpha'$-complete equations is not sensitive to the boundary conditions of the metric perturbations. To understand how this can affect some results, consider perturbing the vacuum Friedmann equations $H^2 = 0 = \dot{H}$, but in terms of the metric perturbation $a(t) = a^{(0)}(t)+a^{(1)}(t,x)$ :
\begin{equation}
    H^2 = H_{(0)}^2 + \frac{2H_{(0)}}{a_{(0)}} \left(\dot{a}_{(1)} - H_{(0)} a_{(1)}\right) = 0.
\end{equation}
Since the background solution is $H_{(0)} = 0$, $\dot{a}_{(1)}$ is not constrained to first order. However, in standard GR, the trace part of the $ij$-component of the perturbed Einstein's equations would give \cite{Chiaffrino:2020akd}
\begin{equation}
    \partial^i \partial_i a_{(1)} =0, 
\end{equation}
and the invertibility of the Laplacian in the support domain of the perturbation would guarantee that $a_1 = 0$. This is equivalent to saying that the perturbation does not have a constant (zero-mode) piece, which would be fixed by a condition on the boundary of the domain. Such a boundary condition is typically assumed to be vanishing at spatial infinity, which is related to the invertibility of the Laplacian. This property is also required for writing the longitudinal vector and divergence-full traceless tensor part of $g_{\mu\nu}^{(1)}$ in terms of the perturbed metric.  

When we do not have any information about the spatial boundary condition on the metric perturbations, such as in the duality invariant setup, we can look at the equation at the next-order in perturbation to fix the zero-mode of first-order perturbations. For instance, the second-order term in the perturbed equation of motion 
\begin{equation}
    0=H^2 = H_{(0)}^2 + \frac{2H_{(0)}}{a_{(0)}} \left(\dot{a}_{(1)} - H_{(0)} a_{(1)}\right)+\frac{\dot{a}_{(1)}^2}{a_{(0)}^2} + \frac{2H_{(0)}}{a_{(0)}}\left(\frac{3H_{(0)} a_{(1)}^2}{a_{(0)}} - \frac{2a_{(1)}\dot{a}_{(1)}}{a_{(0)}} \right) = \frac{\dot{a}_{(1)}^2}{a_{(0)}^2},
\end{equation}
sets the perturbation $a_{(1)}$ to zero. So, ambiguities in the solutions for first-order perturbations in the metric are fixed by higher-order terms in the perturbed equations. 

\section{Conclusion}

In this paper, we have revisited the non-linear stability of Minkowski and dS spacetimes in the context of $\alpha'$-complete cosmology. We have found that solutions with $\dot{\Phi} =0$ are unstable, confirming the results of \cite{Bieniek:2022mrv} for the vacuum dS case. However, we have also taken into account perturbations in the matter sector, and our analysis is free of potential inconsistencies due to the invertibility of the constraint, and loss of information after taking derivatives of the perturbed equations. 
We also compared perturbation theory in the duality invariant setup with standard cosmological perturbations in Einstein gravity. 

In a dynamical-system sense, the results for dS solutions reveal the existence of a saddle-node bifurcation: there are no fixed points for $y$ when $F_0<0$, while $F_0>0$ allows for two fixed points at $y_0 = \pm \sqrt{F_0}$, one of which is stable and the other unstable. At the bifurcation point, when $F_0=0$, there is only one fixed point $(y_0=0)$, which is unstable at the non-linear order.

The results in section \ref{sec:Bardeenvariables} indicate that the restricted approach of considering homogeneous perturbations for $\alpha'$-exact backgrounds is self-consistent. In particular, we have shown that one can define time-reparameterization invariant variables even though vector and tensor perturbation modes were neglected in the homogenous and isotropic analysis. Note, however, that $B_1$ and $B_2$ are not duality invariant. 

As future directions, one might consider developing a gauge \emph{and} duality-invariant formulation of the perturbative analysis, i.e., taking $\mathcal{S}$ as the fundamental field variables and perturbing the action \eqref{eq:hohmzwiebachaction} directly. This would allow studying the stability of solutions that also include the $b$-field, such as the ones discussed in \cite{Nunez:2020hxx, Bernardo:2021xtr}. The homotopy transfer approach put forward in \cite{Chiaffrino:2020akd} might be a systematic way to investigate these issues. 

\section*{Acknowledgments}

We would like to thank Robert Brandenberger for useful discussions, and Tomas Codina, Guilherme Franzmann, Krzysztof Meissner, Allison Pinto and Jerome Quintin for comments on an early version of this paper. H.B. also thanks the Symmetries and Cosmology group at Humboldt University Berlin for discussions about string cosmology. H.B. was supported by the Fonds de recherche du Qu\'ebec (PBEEE/303549) and partially by
funds from the NSERC. J. C. thanks Jan Kwapisz and Przemysław Bieniek for insightful discussion on Minkowski spacetime stability. J. C. is supported by funds from IDUB UW SP-501-D111-20-2004310 grant. V. C. is supported in part by the Fonds de recherche du Qu\'ebec (FRQNT).




\bibliographystyle{bibstyle} 
\bibliography{References}





\end{document}